\documentstyle[12pt]{article}
\begin{document}


\newcounter{subequation}[equation]
\renewcommand{\thesubequation}{\alph{subequation}}
\newcommand{\eqnnum}{{\rm ({\theequation\thesubequation})}}

\newcommand{\subequation}{$$%
\refstepcounter{subequation}}

\newcommand{\Endsubequation}{\eqno \eqnnum $$}

\newcommand{\ods}{\par \vspace{1ex} \par}
\newcommand{\lab}{\label}
\newcommand{\eq}{\refstepcounter{equation}}
\newcommand{\bs}{\subequation}
\newcommand{\ns}{\Endsubequation \subequation}
\newcommand{\es}{\Endsubequation \par \noindent}
\newcommand{\be}{\begin{equation}}
\newcommand{\ee}{\end{equation} \par \noindent}

\renewcommand{\]}{$$%
\par \noindent}

\newcommand{\rf}[1]{(\ref{#1})}
\newcommand{\rff}[2]{(\ref{#1}\ref{#2})}
\newcommand{\rfff}[3]{(\ref{#1}\ref{#2}\ref{#3})}
\newcommand{\rfiv}[4]{(\ref{#1}\ref{#2}\ref{#3}\ref{#4})}
\newcommand{\rfv}[5]{(\ref{#1}\ref{#2}\ref{#3}\ref{#4}\ref{#5})}

\newcommand{\odstep}{\par \vspace{1.5ex} \par}
\newcommand{\cit}{\cite}
\newcommand{\bib}{\bibitem}
\newcommand{\bea}{\[ \begin{array}{c}}
\newcommand{\eea}{\end{array} \]}
\newcommand{\bne}{\begin{eqnarray}}
\newcommand{\ene}{\end{eqnarray}}
\newcommand{\no}{\noindent}
\newcommand{\ind}{\indent}
\newcommand{\where}{\noindent where\ }
\newcommand{\ba}{\begin{array}}
\newcommand{\bac}{\begin{array}{c}}
\newcommand{\bal}{\ba{l}}
\newcommand{\ea}{\end{array}}
\newcommand{\N}{\\[1ex]}
\newcommand{\nn}{\nonumber}

\newcommand{\bm}{\bf}
\newcommand{\bb}[1]{{\bm #1}}
\newcommand{\txt}{\textstyle}
\newcommand{\dis}{\displaystyle}
\newcommand{\Tr}{{\rm Tr}}
\newcommand{\RE}{{\rm Re}}
\newcommand{\IM}{{\rm Im}}
\newcommand{\im}{{\rm im}}
\newcommand{\spa}{{\rm span}}
\newcommand{\E}{{\bm E}}
\newcommand{\R}{{\bm R}}
\newcommand{\C}{{\bm C}}
\newcommand{\G}{{\bm G}}
\newcommand{\h}{{\bm H}}
\newcommand{\GL}{{\bm GL}}
\newcommand{\SL}{{\bm SL}}
\newcommand{\SO}{{\bm SO}}
\newcommand{\so}{{\bm so}}
\newcommand{\SU}{{\bm SU}}
\newcommand{\su}{{\bm su}}
\newcommand{\SP}{{\bm SP}}
\newcommand{\sP}{{\bm sp}}
\newcommand{\Cl}{{\cal C}}
\newcommand{\id}{{\bm I}}
\newcommand{\I}{{\bm i}}
\newcommand{\J}{{\bm j}}
\newcommand{\K}{{\bm k}}
\newcommand{\r}{{\bm r}}
\newcommand{\n}{{\bm n}}
\newcommand{\e}{{\bm e}}
\newcommand{\f}{{\bm f}}
\newcommand{\al}{\alpha}
\newcommand{\bet}{\beta}
\newcommand{\k}{\kappa}
\newcommand{\s}{\sigma}
\newcommand{\g}{\gamma}
\newcommand{\z}{\zeta}
\newcommand{\la}{\lambda}
\newcommand{\ep}{\varepsilon}
\newcommand{\ph}{\varphi}
\newcommand{\th}{\vartheta}
\newcommand{\om}{\omega}
\newcommand{\sh}{\sinh}
\newcommand{\ch}{\cosh}

\newcommand{\dd}{\partial}
\newcommand{\hf}{\frac{1}{2}}
\newcommand{\m}{\lf( \bac}
\newcommand{\mm}{\lf( \ba{cc}}
\newcommand{\mmm}{\lf( \ba{ccc}}
\newcommand{\miv}{\lf( \ba{cccc}}
\newcommand{\ema}{\ea \rg)}

\newcommand{\lf}{\left}
\newcommand{\rg}{\right}
\newcommand{\ov}{\over}
\newcommand{\ovl}{\overline}
\newcommand{\M}[1]{\mid #1\mid}
\newcommand{\tm}{\! \times \!}

\newtheorem{Th}{Theorem}
\newtheorem{prop}{Proposition}[section]
\newtheorem{lem}[prop]{Lemma}
\newtheorem{rem}[prop]{Remark}
\newtheorem{cor}[prop]{Corollary}
\newtheorem{Def}[prop]{Definition}


\title{{\bf Isothermic surfaces in $\E^3$ as soliton surfaces}%
\thanks{The work supported in part by the grants 566/2/91 GR
10 (KBN 2 0168 91 01) and PB 1274/P3/92/02 (KBN 2 2303 91 02).}}

\author{Jan Cie\'sli\'nski \\ {\footnotesize Warsaw University Division in
Bia\l ystok} \\ {\footnotesize Institute of Physics, ul.\ Lipowa 41, 15-424
Bia\l ystok, Poland} \and Piotr Goldstein \\ {\footnotesize Soltan Institute
for Nuclear Studies, ul.  Ho\.za 69, 00-681 Warsaw, Poland} \and Antoni Sym
\\ {\footnotesize Warsaw University} \\ {\footnotesize Institute of
Theoretical Physics, ul.  Ho\.za 69, 00-681 Warsaw, Poland}}

\maketitle

\begin{abstract}
We show that the theory of isothermic surfaces in $\E^3$ -- one of the
oldest branches of differential geometry -- can be reformulated within the
modern theory of completely integrable (soliton) systems.  This enables one
to study the geometry of isothermic surfaces in $\E^3$ by means of powerful
spectral methods available in the soliton theory.  Also the associated
non-linear system is interesting in itself since it displays some
unconventional soliton features and, physically, could be applied in the
theory of infinitesimal deformations of membranes.
\end{abstract}
\newpage
\section{Introduction}

There is no doubt today that some fundamental ideas and many concrete
results of the modern theory of completely integrable (soliton) systems
can be traced back over a century 
to
the classical differential geometry. For instance, one of the most important
ingredients of the soliton theory is the theory of B\"acklund
transformations. The first example of the B\"acklund transformation (for
the celebrated sine-Gordon eq.) originated in the works of the great
differential geometers of XIX century (G.Darboux and L.Bianchi) to be
finally formulated by A.V.B\"acklund in 1880.

In general, it turns out that a careful study of some other works by
G.Darboux and, notably, by L.Bianchi leads to a conclusion that some their
results are of a genuine soliton nature.  Exactly in this way (by studying
the paper \cit{Bi1} by L.Bianchi) in 1991 we realized that, presumably, the
theory of the so called isothermic surfaces -- one of the oldest branches of
differential geometry \cit{Voss} which recently is a subject of some modern
studies \cit{Chern,Svec,Palmer,Bob} -- can be reformulated within the modern
approach of {\em soliton surfaces} \cit{Sym}.  This conjecture has been
further confirmed by different tests.

Based on results of this paper (presented in the short report
\cit{CGS-short}) Pinkall's Berlin Group have written a number of preprints
including \cit{BHPP,BP}.

\section{Isothermic surfaces and Bonnet surfaces}

Consider an arbitrary immersed surface $S$ in $\E^3$ without umbilic points
(in umbilic points both principal curvatures coincide).
The question is: What conditions one should impose on $S$ to guarantee the
existence of {\em infinitesimal} isometric deformations preserving principal
curvatures (or, equivalently, the mean curvature) invariant ?

The answer -- well known to the geometers of XIX century -- is that
 the curvature coordinates, say $u$ and $v$, are conformal (after a proper
reparameterization), i.e. the fundamental forms read

\be   \lab{I,II}
\ba{ll}
 I  = e^{2\th}( du^2 + dv^2 ) & \qquad ({\rm metric}) \ , \N
 II = e^{2\th}(k_2 du^2 + k_1 dv^2) & \qquad ({\rm 2\mbox{-}nd\
 fundamental\ form})\ ,
\ea \ee

\where $k_1$=$k_1(u,v)$ and $k_2$=$k_2(u,v)$ are principal curvatures and
$\th$=$\th(u,v)$.

 The triplet  $(\th,k_1,k_2)$ has  to satisfy the following system  of
non-linear  partial  differential eqs. (Gauss-Mainardi-Codazzi eqs.):

\eq      \lab{GMC}
\bs      \lab{GMCa}
      \th,_{uu} + \th,_{vv} + k_1 k_2 e^{2\th}  =  0  \ ,  \N
\ns      \lab{GMCb}
       k_1,_u + (k_1 - k_2)\th,_u   =   0  \ , \N
\ns      \lab{GMCc}
                 k_2,_v + (k_2 - k_1)\th,_v   =   0  \ ,
\es

\where comma denotes differentiation.

And vice versa, any solution $(\th,k_1,k_2)$ defines uniquely some
isothermic surface in $\E^3$ modulo a rigid motion in $\E^3$.  In other
words, the geometry of isothermic surfaces in $\E^3$ without umbilic points
is completely encoded in the system \rf{GMC}.  The theory of isothermic
surfaces at umbilics is much more difficult.

\ods

Surfaces in $\E^3$ admitting global and non-trivial (rigid motions are
excluded) isometries preserving principal curvatures (or, equivalemtly, the
mean curvature $H := \hf (k_1+k_2)$) are called Bonnet surfaces
\cit{Chern,Palmer,Bob} after O.Bonnet who was the first to study such
surfaces in 1867.

Bonnet surfaces constitute a proper subset of isothermic surfaces.  Indeed,
the totality of isothermic surfaces can be parameterized by 4 functions of a
single variable each (the order of the system \rf{GMC} is 4 !).  For a more
detailed proof see \cit{Svec}.  On the other hand S.S.Chern \cit{Chern} has
shown that the class of Bonnet surfaces consists of two subclasses:  i)
surfaces of $H$=const (to select a surface of $H$=const one needs 2
functions of a single variable) and ii) some family of surfaces
parameterized by 6 parameters.

\ods

In 1903 P.Calapso \cit{Calapso} on performing a series of remarkable
transformations was able to replace the system \rf{GMC} by the following
single equation of 4-th order

\be  \lab{Cal}
        \lf( \frac{w,_{uv}}{w} \rg),_{uu} +
        \lf( \frac{w,_{uv}}{w} \rg),_{vv} + \lf( w^2 \rg) ,_{uv} = 0 \ .
\ee

Finally, we mention an interesting physical application of isothermic
(in particular: Bonnet) surfaces \cit{Sabitov}. Imagine
an elastic membrane moving
isometrically in $\E^3$. We require that the difference of pressures
between both sides of the membrane is a constant of motion at each point
(though it may vary from point to point). One can show that such a
motion is admitted for Bonnet surfaces only. Obviously, infinitesimally
short evolution corresponds to isothermic surfaces.

\section{The starting point of the research}

In the paper \cit{Bi1} (page 98) one can find the following formulae:
\be   \lab{teneqs}
\bal
   \la,_u = -\th,_v \mu - k_2 e^\th \om + m \s e^\th + m e^{-\th} \ph\ , \N
   \mu,_u = \th,_v \la\ , \qquad \ph,_u = e^\th \la\ , \qquad
   \om,_u = k_2 e^\th \la\ ,  \qquad    \s,_u = e^{-\th}\la\ ,   \N
   \mu,_v = -\th,_u \la - k_1 e^\th \om + m \s e^\th - m e^{-\th} \ph\ ,  \N
   \la,_v = \th,_u \mu\ ,   \qquad    \om,_v = k_1 e^\th \mu\ ,   \qquad
   \s,_v = - e^{-\th} \mu\ ,   \qquad     \ph,_v = e^\th \mu\ ,
\ea
\ee
\no where $\th, k_1$ and $k_2$ is some but fixed solution to the system
\rf{GMC}, $m$ is a non-zero parameter, and $\la, \mu, \om, \ph$ and $\s$ are
5 unknowns.

The important point is that the integrability conditions for the system
\rf{teneqs} are identical with the system \rf{GMC} and, moreover, the system
\rf{teneqs} contains a free parameter, namely: $m$.

In the soliton theory such a coexistence of the linear system ({\it linear
problem}) containing a parameter ({\it spectral parameter}) with the
corresponding non-linear system strongly suggests (by no means proves !)
that the non-linear system could be integrable in the sense of the soliton
theory.

The  other important fact is that the quadratic form
\be  \la^2 + \mu^2 + \om^2 - 2 m \ph \s   \lab{kwadrat}  \ee
\no does not depend on $u$ and $v$.

One could name the linear problem \rf{teneqs} for the system \rf{GMC} -- a
Darboux-Bianchi linear problem since it was introduced by G.Darboux and
later on became a subject of intensive studies by L.Bianchi and his Pisa
school.

Solving the system \rf{teneqs} one is able to obtain a new solution
$(\th',\,k'_1,\, k'_2)$ of the system \rf{GMC}:

\eq    \lab{CDB}
\bs
                e^{\th'} = \pm \frac{\ph}{\s} e^{-\th} \ ,     \lab{theta}
\ns
        k_1' e^{\th'} = \pm \lf( k_1 e^{\th} + \frac{\om}{\ph} e^{\th}
                        - \frac{\om}{\s} e^{-\th} \rg) \ ,      \lab{k1}
\ns
      k_2' e^{\th'} = \mp \lf( k_2 e^{\th} + \frac{\om}{\ph} e^{\th}
                        + \frac{\om}{\s} e^{-\th} \rg) \ ,      \lab{k2}
\es
\where the upper sign (lower sign) corresponds to $m>0$ ($m<0$).

One can call \rf{CDB} which is a kind of B\"acklund transformation a
 ``classical Darboux-Bianchi transformation''.  In fact in \cit{Bi2}
 L.Bianchi proved the abelian property of the transformation \rf{CDB} which
 is a characteristic feature of B\"acklund transformations.

The linear problem \rf{teneqs} and the classical Darboux-B\"acklund
transformation were the starting point of our
research of the subject which we undertook in 1991.

\section{\hspace{-7pt plus 7pt} Painlev\'e analysis of the nonlinear systems
associated with isothermic surfaces} \lab{PG}

First of all we applied a relatively simple test of the integrability -- the
so called ``Painlev\'e test'' \cit{ARS,Conte} to the systems \rf{GMC} and
\rf{Cal}.

The Painlev\'e test is performed in its classical form \cit{ARS} extended to
partial differential equations in \cit{WTC}, i.e. by assuming a
solution in the form of a Laurent series about an arbitrary
singularity manifold $\Phi(u,v)=0$ and checking compatibility of
the resulting recurrence formulae.
Detailed discussion of the  meaning, validity and
techniques of this test may be found in \cit{Conte}.
The test is
carried out for system \rf{GMC} and for the Calapso equation
\rf{Cal}. Both the GMC system and the Calapso equation pass the
test. For the system \rf{GMC} we also find the B\"acklund
transformation.

Equations \rf{GMC} are cast into a polynomial form by substitution
\be  \lab{subKM}
k_1 = K \exp(-\th),~~~~~k_2 = M \exp(-\th)
\ee
The Laurent expansion of $\th$ is supplemented by a logarithmic
term due to potential character of this variable \cit{Conte}. Variables
$K$ and $M $ begin their series with $\Phi^{-1}$.
The resonances (sometimes
called 'indices' \cit{CFP}) arise at terms of number $r=0$ and $r=2$ in the
Laurent expansions of $\th,~ K,~ M$ (the logarithmic term is not
numbered). The resonance at $r=2$ is double. All of them are compatible.

Truncation of these Laurent series on terms od order $\Phi^0$ yields a
B\"acklund transformation. The transformation between
 $\th,~K,~M$ and $\th_0,~K_1,~M_1$ reads
\eq \lab{Backl}
\bs  \lab{Backla}
 \th = \pm \ln(\Phi) + \th_0,~~~~K=iS/\Phi + K_1,~~~~M = \mp iS/\Phi + M_1,
\ns \lab{condPhi}
        \Phi,_v \th_0,_u + \Phi,_u \th_0,_v \pm \Phi,_{uv} = 0,
\ns \lab{condM1}
\pm\*(i \Phi,_u M_1 + S \th_0,_u) + S,_u = 0,
\ns \lab{condK1}
i \Phi,_v K_1  \mp S \th_0,_v - S,_v = 0~~,~~{\rm where}
\ns \lab{S}
            S = ({\Phi,_u}^2 + {\Phi,_v}^2)^{1/2}
\es
The first three of these equations \rff{Backl}{Backla}
are truncated expansions of $\th,~K,~M$
(the indices in $\th_0,~K_1,~M_1$ correspond to the numbering of
terms in the Laurent series).
Condition \rff{Backl}{condPhi} may easily be recognized as vanishing of
the next term $\th_1$ in the Laurent series of the potential $\th$
while the last two conditions, \rff{Backl}{condM1} and
\rff{Backl}{condK1} are recurrence relations for $M_1$ and $K_1$,
respectively, when $\th_1$ vanishes.

Compatibility conditions for this overdetermined system are indeed
equations \rf{GMC}. Namely,  \rff{GMC}{GMCb} and \rff{GMC}{GMCc}
(substituted according to \rf{subKM}) ensure compatibility
$S,_{uv} = S,_{vu}$ of \rff{Backl}{condM1} and
\rff{Backl}{condK1}. We were also able to obtain
 equation \rff{GMC}{GMCa} as a fairly complicated combination of
compatibility conditions for \rf{Backl} (useful hints for construction
of that combination are provided by the fact that second order
coefficients in the Laurent expansions of $\th,~K,~M$ should vanish).

The B\"acklund transformation \rf{Backl} is different from the
transformation \rf{teneqs}, \rf{CDB}; when applied to a solution which
is real for real $(u,\,v)$, it takes at least one of the dependent
variables out of the real axis.

The Calapso equation is given a polynomial form, by multiplcation of
both hand sides of \rf{Cal} by $w^3$. The Laurent series for $w$ begins
with a term of order $\Phi^{-1}$. Compatibility conditions arise at
$r=2$, $r=3$ and $r=4$ (one condition per resonance). The check for
compatibility is a bit cumbersome but straightforward. The conditions
are satisfied at all the resonances.

\section{Further developments}

It is not difficult to notice that the invariant quadratic form \rf{kwadrat}
is of the signature $(++++-)$.  This enables one to rewrite the
Darboux-Bianchi linear problem as an $\so(4,1)$-linear problem (here we use
the standard terminology of the soliton theory).  Namely, assuming $m>0$ we
perform the following transformation

\be
    \tilde{\ph} = \sqrt{\frac{m}{2}} \ (\s - \ph) \ ,\qquad {\rm and} \qquad
    \tilde{\s} = \sqrt{m \ov 2} \ (\s + \ph) \ ,
\ee

\no and then the Darboux-Bianchi linear problem \rf{teneqs} can be rewritten
as follows

\be   \lab{SOproblem}
\bal
        \psi,_u = \lf( - \th,_v \f_{12} - k_2 e^\th \f_{13} +
        \z \sh\th\ \f_{14}\ +\  \z \ch\th\ \f_{15} \rg) \psi \ , \N
        \psi,_v = \lf( \th,_u \f_{12} - k_1 e^\th \f_{23} +
        \z \ch\th\ \f_{24}      \ +\    \z \sh\th\ \f_{25} \rg) \psi \ ,
\ea
\ee

\where we put $\psi:=(\la,\mu,\om,\tilde{\ph},\tilde{\s})^T$ and
$\z:=\sqrt{2m}$, the latter as a ``spectral parameter'', and, finally the
matrices $\f_{ij}\ (1\leq i<j\leq 5)$ constitute the standard basis of the
Lie algebra $\so(4,1)$:
\be
\bal
  (\f_{ij})_{\al\bet} = \delta_{i\al} \delta_{j\bet} - \delta_{i\bet}
\delta_{j\al} \qquad   ({\rm for} \ i<5,\, j<5 ),  \N
  (\f_{i5})_{\al\bet} =  \delta_{i\al} \delta_{5\bet} + \delta_{i\bet}
\delta_{5\al} \ ,
\ea
\ee

\where $1\leq \al,\bet \leq 4$ and $\delta_{jk}$ is Kronecker's delta.

Certainly, the integrability conditions for \rf{SOproblem} are still the
same, i.e.  they are given by the system \rf{GMC}.

\ods

There are some obvious disadvantages of the linear problem \rf{SOproblem}:
too many zero-entries in the matrices of the problem and the large dimension
of the matrices.  In this context a natural question arises:  {\em what is a
minimal Lie algebra containing both matrices of the linear problem}
\rf{SOproblem} for an arbitrary choice of $\th,k_1,k_2$ and $\z \in \R$ ?
This problem turned out to be non-trivial.  It took a few months to find the
answer \cit{dim=6}.

	The obtained minimal Lie algebra is the Lie algebra of the rigid
motions in $\E^3$ (semidirect sum of the Lie algebra of rotations and
translations).  It was a discouraging result:  usually soliton systems are
related to semi-simple Lie algebras.

  The only way out is to make use of the well known isomorphism between
$\so(4,1)$ and $\sP(1,1)$ \cit{izomorfizm}.  For instance, this isomorphism
can be given by

\be
                \so(4,1) \ni \f_{jk} \quad \mapsto \quad
                \hf \e_{jk} := \hf \e_j \e_k \in \sP(1,1) \ .\lab{izomor}
\ee

\no where complex $4\tm 4$ matrices $\e_j$ $(j=1,\ldots,5)$ are defined as
follows
\be  \lab{ej}
\bac
        \ba{ccc}
     \e_1 = \mm  0 & i\s_2 \\  -i\s_2 & 0  \ema\ ,  &
     \e_2 = \mm -\s_1 & 0 \\ 0 & -\s_1 \ema\ , &
     \e_3 = \mm -\s_2 & 0 \\ 0 & \s_2 \ema\ ,
        \ea                                                \\[4ex]
   \ba{cc}
     \e_4 = \mm  0 & \s_2 \\ \s_2 & 0 \ema\  , &
     \e_5 = \mm  i\s_3 & 0 \\ 0 & i\s_3 \ema .
         \ea
\ea
\ee

\no and $\ \s_k \ (k=1,2,3)\ $ are standard Pauli matrices:

\be \lab{Pauli}
\ba{ccc}
\s_1 = \mm 0 & 1 \\ 1 & 0 \ema\ , & \s_2 = \mm 0 & -i \\
i & 0 \ema\ , & \s_3 = \mm 1 & 0 \\ 0 & -1 \ema\ .
\ea \ee
One can check in a straightforward way the following properties
\be
\bal
\e_k \e_j = - \e_j \e_k \qquad (k \neq j) \ , \N
\e_1^2 = \e_2^2 = \e_3^2 = \e_4^2 = -\e_5^2 = \id \ , \N
i\e_1\e_2\e_3\e_4\e_5 = \id
\ea \ee
\no ($\id$ is an
identity matrix) which mean that $i\e_1,\ldots,i\e_5$ generate an algebra
isomorphic to the subalgebra of even elements of the Clifford algebra
$\Cl(1,4)$.

The isomorphism \rf{izomor} enables one to rewrite the $\so(4,1)$-linear
problem
\rf{SOproblem} as the following $\sP(1,1)$-linear problem
\eq  \lab{SPproblem}
\bs  \lab{U}
    \Psi,_u = \hf\,\e_1 \lf( -  \th,_v \e_2  - k_2 e^\th \e_3 +
        \z \sh\th\ \e_4  + \z \ch\th\ \e_5 \rg) \Psi \ ,
\ns     \lab{V}
          \Psi,_v = \hf\, \e_2 \lf( - \, \th,_u \e_1  - k_1 e^\th \e_3
                 + \z \ch\th \e_4      +      \z \sh\th\ \e_5 \rg) \Psi \ ,
\es

\no where $\Psi=\Psi(u,v;\z)$ is a non-degenerate complex $4 \tm 4$ matrix.

   We conclude this section mentioning that in 1992 we asked our colleagues
(Ruud Martini's group) of the Math.  Dep.  of the University of Twente (The
Netherlands) to apply their original technique (based on symmetries) to give
an independent proof of the integrability of the underlying non-linear
system \rf{GMC}.  Indeed, in 1993 Theo van Bemmelen \cit{TvB-diss} found the
so called recursion operator for the system \rf{GMC} providing us with yet
another proof of the soliton nature of the system.

\section{Isothermic surfaces as soliton surfaces}

The integrability of the system \rf{GMC} which is an implicit description of
the isothermic surfaces in $\E^3$ means that the class of all (local)
isothermic surfaces is yet another example of integrable geometry.  This
subject can be studied within the approach of soliton surfaces \cit{Sym}.

  We recall that the fundamental forms (e.g.  those of isothermic surfaces
in $\E^3$:  \rf{I,II}) define a surface in $\E^3$ uniquely (modulo rigid
motion in $\E^3$).  As a rule, it is very difficult to recover the explicit
expression for the position vector to the surface from the knowledge of its
fundamental forms.

Fortunately, when one deals with the integrable geometry of surfaces
(submanifolds), the problem of reconstruction of a surface (submanifold) can
be simplified greatly.  Namely, it is the existence of the associated
$\z$-dependent linear problem (e.g.  \rf{SOproblem} or \rf{SPproblem} for
isothermic surfaces in $\E^3$) for the ``wave function'' $\Psi$.  One can
prove \cit{Sym,Sym'} that the formula

\be
        R = \Psi^{-1} \Psi,_{\z} |_{\z=\z_0}        \lab{Sym}
\ee

\no defines a class of surfaces (submanifolds) immersed into the associated
Lie algebra of the linear problem with exactly the same underlying nonlinear
system as the one of the initial integrable geometry.  In many cases
(\cit{Bob}, \cit{Sym} and references quoted therein) the formula \rf{Sym}
reconstructs our initial integrable geometry.

It is interesting that in the case of the integrable geometry of isothermic
surfaces in $\E^3$ the formula \rf{Sym} requires some modification.  Indeed,
one can show that in this case the formula \rf{Sym} ($\z_0=0$) defines class
of surfaces immersed in a 6-dim.  linear subspace of the Lie algebra
$\sP(1,1)$ spanned by $\e_k \e_4$, $\e_k \e_5$ $(k=1,2,3)$.

\ods
The main results of the paper reads (for proof see \cite{DB-izot}):
\ods

{\it  Given a solution  $(\th,k_1,k_2)$ to the system \rf{GMC},
 the corresponding isothermic surface may be recovered  as follows.
\renewcommand{\theenumi}{(\alph{enumi})}
\begin{enumerate}
\item
Insert the solution $(\th,k_1,k_2)$ into matrices of the linear problem
                \rf{SPproblem}.

\item  \lab{difficult}
 Compute the corresponding wave function $\Psi=\Psi(u,v;\z)$.

\item  \lab{projection}
Compute
\be
       \r= P \Psi^{-1}\Psi,_{\z}|_{\z=0}    \lab{Sym'}
\ee
where $P$ is the constant projector given by $P:= \hf (1-\e_{45})$.
\item Decompose $\r$ in the basis
$\f_k := \frac{1}{4} \e_k (\e_4 + \e_5)$,  $k=1,2,3$:
\be
			\r = X \f_1 + Y \f_2 + Z \f_3 \ .
\ee
\end{enumerate}
The map
$ \R^2 \ni (u,v) \mapsto (X,Y,Z) \in \R^3$ describes explicitly
the surface we look for. }
\ods
Performing in the step~\ref{projection} the projection
$I-P$ instead of $P$ we obtain the so called dual surface, or
Christoffel transform  of\  $\r$. This surface is isothermic as well
and its fundamental forms are parameterized by
$\th'=-\th$, $k_1'=e^{2\th} k_1$, $k_2'=-e^{2\th} k_2$ (compare  \cit{Bi1}).
\ods

In the above algorithm the step~\ref{difficult} is certainly the most
difficult.  However, if the triplet $(\th,k_1,k_2)$ is the $N$-soliton
solution, the formalism of the soliton theory (e.g.  \cit{N-soliton})
enables one to compute the corresponding $\Psi$ explicitly (see
\cite{DB-izot,DBT-JMP}).  In this way one arrives at the expression for
1-soliton isothermic surface

\be \lab{1-sol}
  \r_1 = \m u \\ v \\ 0 \ema +
  \frac{2}{\ch v \ch\g - \cos u}
        \m  \sin u \\  - \sh v  \ch\g  \\  - \sh\g \ema
\ee

\no where $\g$ is a constant parameter. A sample of such surfaces
is shown on Fig.~1.

We conclude with the statement that within the approach of soliton surfaces
one can reconstruct and generalize all the classical findings by G.Dar\-boux
and L.Bian\-chi \cit{Bi1}.  In particular one can derive (by the standard
dressing method) the classical Darboux-B\"acklund transformation \rf{CDB}.
The detailed discussion of these results is given in \cit{DB-izot}.

\section*{Acknowledgements}
Special thanks are due to our colleagues and friends:  Peter
Gragert, Reinhard Meinel and Theo van Bemmelen, for their interests,
comments and useful hints.

\vspace{1.5cm}

\section*{Figure caption}

Fig. 1. One-soliton isothermic surfaces \rf{1-sol} for
three values of the parameter
$\g$. In the limit of high $\g$ the surface becomes a
self-intersecting pipe whose cross section has a shape of
the Wadati-Konno-Ichikawa loop soliton \cit{WKI}.

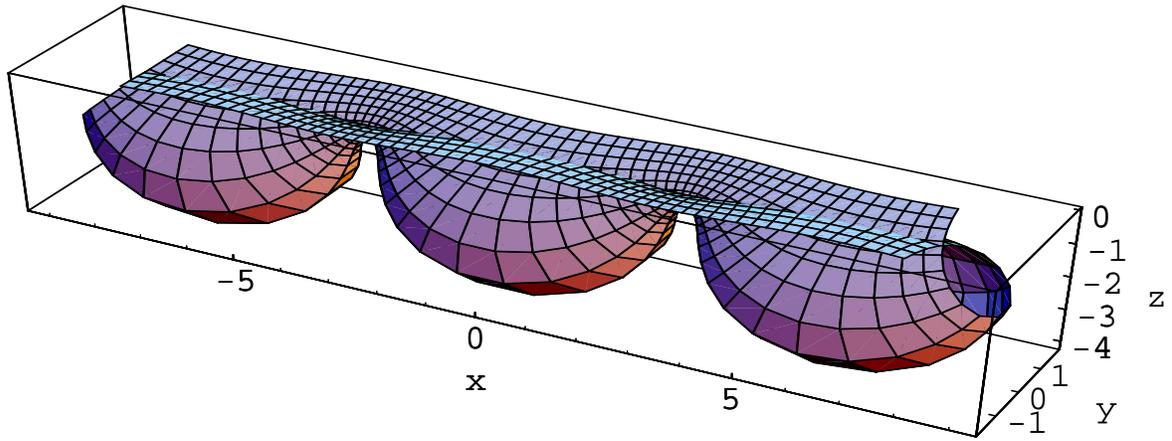
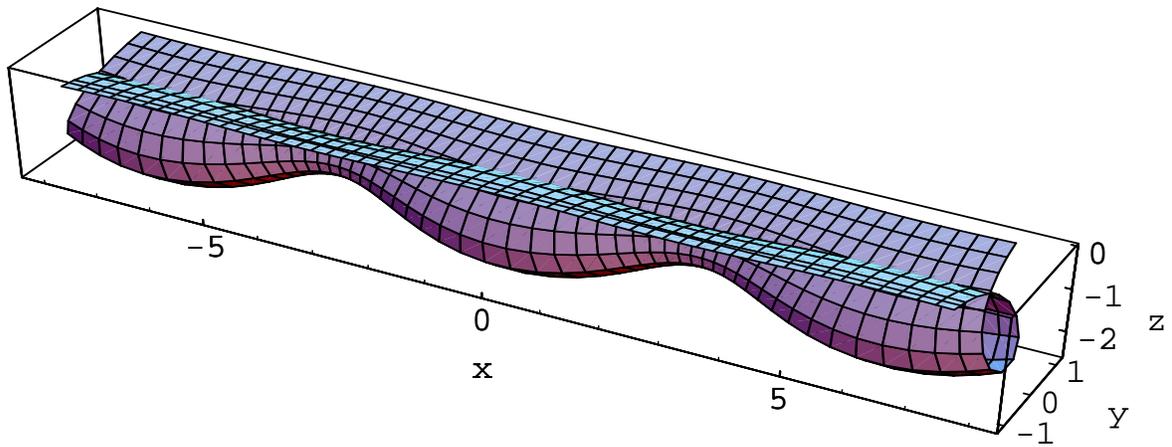
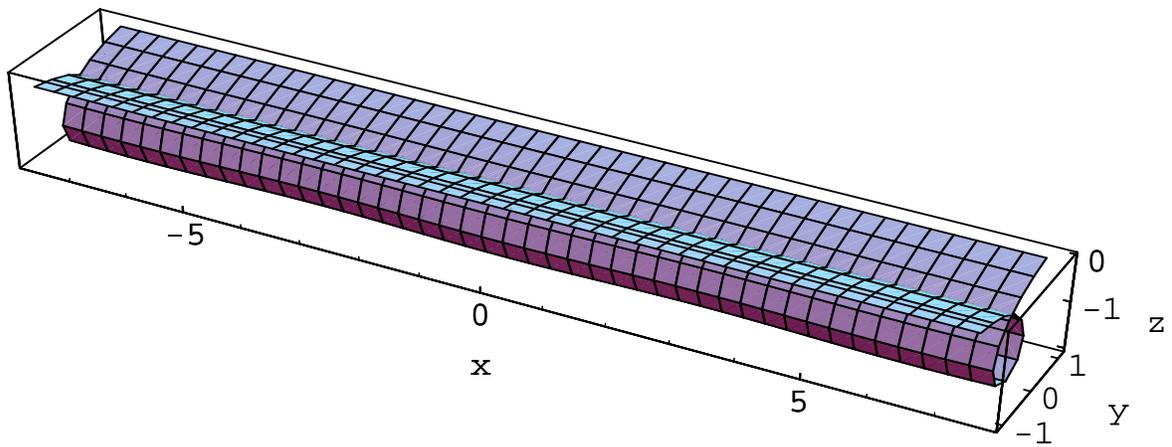
\end{document}